\documentclass[sigconf]{acmart}
\AtBeginDocument{%
  }

\copyrightyear{2026}
\acmYear{2026}
\setcopyright{cc}
\setcctype{by}
\acmConference[RecSys '26]{20th ACM Conference on Recommender Systems}{September 27-October 02, 2026}{Minneapolis, MN, USA}
\acmBooktitle{20th ACM Conference on Recommender Systems (RecSys '26), September 27-October 02, 2026, Minneapolis, MN, USA}
\acmDOI{10.1145/3773078.3831858}
\acmISBN{979-8-4007-2284-4/2026/09}

\usepackage{graphicx}
\usepackage{pifont}
\usepackage{url}
\usepackage{xcolor}
\usepackage{xspace}
\usepackage{booktabs}
\usepackage{listings}

\begin{document}
\definecolor{DarkGreen}{RGB}{30,130,30}
\newcommand{\cmark}{\textcolor{DarkGreen}{\ding{51}}}
\newcommand{\xmark}{\textcolor{red}{\ding{55}}}%

\definecolor{DarkYellow}{RGB}{255,204,0}
\newcommand{\ytriangle}{\textcolor{DarkYellow}{\LARGE$\triangle$}}

\newcommand{\proposed}{\textsc{WatchLens}\xspace}
\newcommand{\smallsection}[1]{{\vspace{0.05in} \noindent \bf {#1.\hspace{5pt}}}}

\newcommand{\red}[1]{\textcolor{red}{#1}}
\newcommand{\blue}[1]{\textcolor{blue}{#1}}
\title{\textsc{WatchLens}: A Configurable Platform for Online Video Recommendation Experiments}

\author{Deogyong Kim}
\orcid{0009-0006-4497-2997}
\affiliation{%
  \department{Department of Artificial Intelligence}
  \institution{Yonsei University}
  \city{Seoul}
  \country{Republic of Korea}
  }
\email{legenduck@yonsei.ac.kr}

\author{Dongha Lee}
\orcid{0000-0003-2173-3476}
\authornote{Corresponding author}
\affiliation{%
  \department{Department of Artificial Intelligence}
  \institution{Yonsei University}
  \city{Seoul}
  \country{Republic of Korea}
    }
\email{donalee@yonsei.ac.kr}

\renewcommand{\shortauthors}{Kim and Lee}

\begin{abstract}
Studying how video recommender systems shape user behavior requires online experiments that link playback behavior with the recommendation conditions that produced it. 
Existing user-study infrastructure provides one or the other, but not both within a single experimentation workflow. We present \proposed, an open-source platform that fills this gap. 
\proposed adopts a modular architecture in which user interfaces, content sources, and recommendation policies are independently configurable, with policies assignable separately to the feed and the watch page, while a standardized logging layer attaches the recommendation policy and ranking position to every event at recording time.
This design enables researchers to analyze how recommendation policies and ranking positions shape downstream playback behavior, session continuation, and navigation between the feed and the watch page, with the linkage between policy and outcome available in each event rather than reconstructed afterwards. 
We demonstrate \proposed with a short-form video case study that holds the interface, feed policy, and content pool constant while varying only the watch-page policy, showing how the platform supports session-level comparison of recommendation effects on real viewing behavior. 
\proposed is released as a publicly available, single-server deployable system for reproducible online video recommendation research.
\end{abstract}




\maketitle

\section{Introduction}
\label{sec:intro}

Offline datasets and benchmarks have served as a foundation for recommender systems research~\cite{zangerle2022evaluating}, enabling reproducible evaluation and comparison across models. 
However, such data are observational records collected under fixed historical conditions and cannot reveal how user behavior would change under new policies or experimental setups~\cite{schnabel2016recommendations}; addressing such questions requires online experimentation with real users~\cite{garcin2014offline, jannach2019measuring}, where recommendation conditions can be controlled and their effects observed. 
This need is particularly pronounced in video recommendation, where user responses extend beyond clicks or item selections into playback- and session-level behaviors such as watch duration, abandonment, skipping, and transitions between videos~\cite{covington2016deep, zhao2023uncovering, pan2023understanding, zhao2019recommending}.

Conducting such experiments, however, imposes a substantial infrastructure burden. 
Before investigating their research questions, researchers must prepare an end-to-end experimentation pipeline spanning user interfaces, content organization, participant and condition assignment, recommendation policy integration, exposure and playback event logging, and data export for analysis. 
This burden raises the barrier to entry for online experimentation, and is acute in video recommendation, where video serving and playback event handling add further engineering requirements. As a result, many studies rely on one-off prototypes built for a single experiment~\cite{heitz2022benefits, jin2018effects, deldjoo2024fairness, wardatzky2025whom}, and experimental components become tightly coupled within each system, making it difficult to vary a single factor in isolation or to reuse the same setup across studies.

Existing resources partially alleviate this burden but do not fully meet the requirements of this experimental setting.
Public video recommendation datasets provide large-scale logs and shared benchmarks~\cite{gao2022kuairec, gao2022kuairand, yuan2022tenrec}, yet because they are collected under fixed historical conditions, they are not suited for experiments in which researchers manipulate recommendation conditions and observe their effects. 
Meanwhile, existing recommender system user-study platforms are useful for delivering content to real users and recording their interactions~\cite{heitz2024informfully, burke2025conducting}, but their logging focuses on modality-independent interactions such as item access, dwell time, and user ratings~\cite{yi2014beyond}, rather than providing a video-native experimentation workflow that links the temporal structure of video playback with recommendation exposure context. 
Consequently, a shared experimentation infrastructure that combines flexible manipulation of recommendation conditions with video-native behavioral observation has yet to be established for online video recommendation research.

To address this gap, we present \proposed, \footnote{Platform: \url{https://github.com/WatchLens/WatchLens} (MIT License). \\ 
Documentation: \url{https://watchlens.github.io}.} an open-source video-native experimentation platform for configuring, deploying, logging, and analyzing online video recommendation experiments. 
\proposed adopts a modular architecture that links user interfaces, recommendation policies, and content sources through defined integration specifications, allowing each component to be independently replaced or extended. 
For every video impression, \proposed records the page on which the recommendation was made, the applied policy, and the ranking position, and associates this information with time-stamped playback events such as play, pause, seek, and watch end. 
From these logs, it computes video-native metrics such as watch duration, watch ratio, and session length~\cite{zhan2022deconfounding, zhao2023uncovering, lin2023tree, xue2022resact}, and exports them in analysis-ready formats.

To demonstrate the practical applicability of \proposed, we conduct a user study in a short-form video browsing environment. 
With the UI and content pool held constant, we vary only the recommendation policy and compare how each user's playback behavior differs across policies. 
This case study shows that \proposed can be used to configure a study tailored to a specific research question, collect real users' playback behavior, and conduct session-level, exposure-aware comparative analysis across study conditions.

The contributions of this paper are as follows:
\vspace{-3pt}
\begin{itemize}
    \item We propose \proposed, a modular platform for online video recommendation experiments that integrates exposure-aware playback logging with independently configurable interfaces, recommendation policies, and content sources.
    \item We release \proposed as a publicly available open-source system, allowing researchers to configure, run, and analyze such experiments within a single deployable environment.
    \item We demonstrate the platform through a user study in a short-form video setting, comparing recommendation policies independently assigned to the feed and the watch page. 
\end{itemize}

\section{Related Work}
\label{sec:relwork}
Video recommendation has been an active research area, with extensive work on watch-time prediction~\cite{covington2016deep, zhao2019recommending, zhan2022deconfounding, lin2023tree, zhao2024counteracting}, sequential video recommendation~\cite{hidasi2015session, kang2018self, sun2019bert4rec, pan2023understanding}, and short-form video recommendation~\cite{zheng2022dvr, cai2023two, cai2023reinforcing, liu2024kuaiformer}. 
Playback-based signals such as watch duration, abandonment, and within-session navigation have emerged as primary indicators of recommendation quality in video settings~\cite{yi2014beyond, zhao2023uncovering}. 
While industrial teams at YouTube, Kuaishou, and TikTok routinely conduct online experiments on their proprietary platforms~\cite{chen2019top, liu2022monolith}, academic research without such infrastructure relies almost exclusively on offline benchmarks, and online video recommendation experiments by external researchers remain scarce.

To support such offline evaluation, a number of public datasets with playback signals have been released. KuaiRec~\cite{gao2022kuairec} provides a near-fully-observed user–item interaction matrix, KuaiRand~\cite{gao2022kuairand} offers logs with random exposure, RecFlow~\cite{liu2025recflow} captures the full multi-stage recommendation pipeline, and VK-LSVD~\cite{poslavsky2026vk} provides industrial-scale data spanning tens of millions of users. 
These datasets have become standard benchmarks for offline model comparison. However, all are collected under fixed historical recommendation policies and interfaces, and therefore cannot support controlled experiments in which researchers manipulate recommendation conditions and observe their effects on real user behavior.

Platforms for conducting recommender system user studies have also been developed. 
EasyStudy~\cite{dokoupil2023easystudy} provides a modular framework for rapidly deploying web-based studies on static items such as movies and books. 
POPROX~\cite{burke2025conducting} provides a hosted infrastructure for news recommendation studies with a managed participant pool. 
Informfully~\cite{heitz2024informfully} supports user studies across text, image, audio, and video modalities within a unified item-feed interface, with video support introduced incrementally to accommodate specific studies.

These platforms have established a foundation for content delivery, interaction tracking, and survey administration in user studies. 
Among them, only Informfully~\cite{heitz2024informfully} accommodates video, but it records playback actions separately from the recommendation exposure that produced them, so the linkage must be reconstructed through post-hoc joining, which risks misattribution when recommendation lists are updated between exposure and interaction. 
It also offers a fixed user interface and does not support assigning separate recommendation policies to the feed and the watch page.
\proposed is instead built around these requirements, treating exposure-aware playback logging, configurable video interfaces, and surface-decoupled policy assignment as first-class capabilities of a platform purpose-built for video recommendation experiments.

\section{\proposed}
\label{sec:method}
\begin{figure}[!t]
\centering
\includegraphics[width=0.89\linewidth]{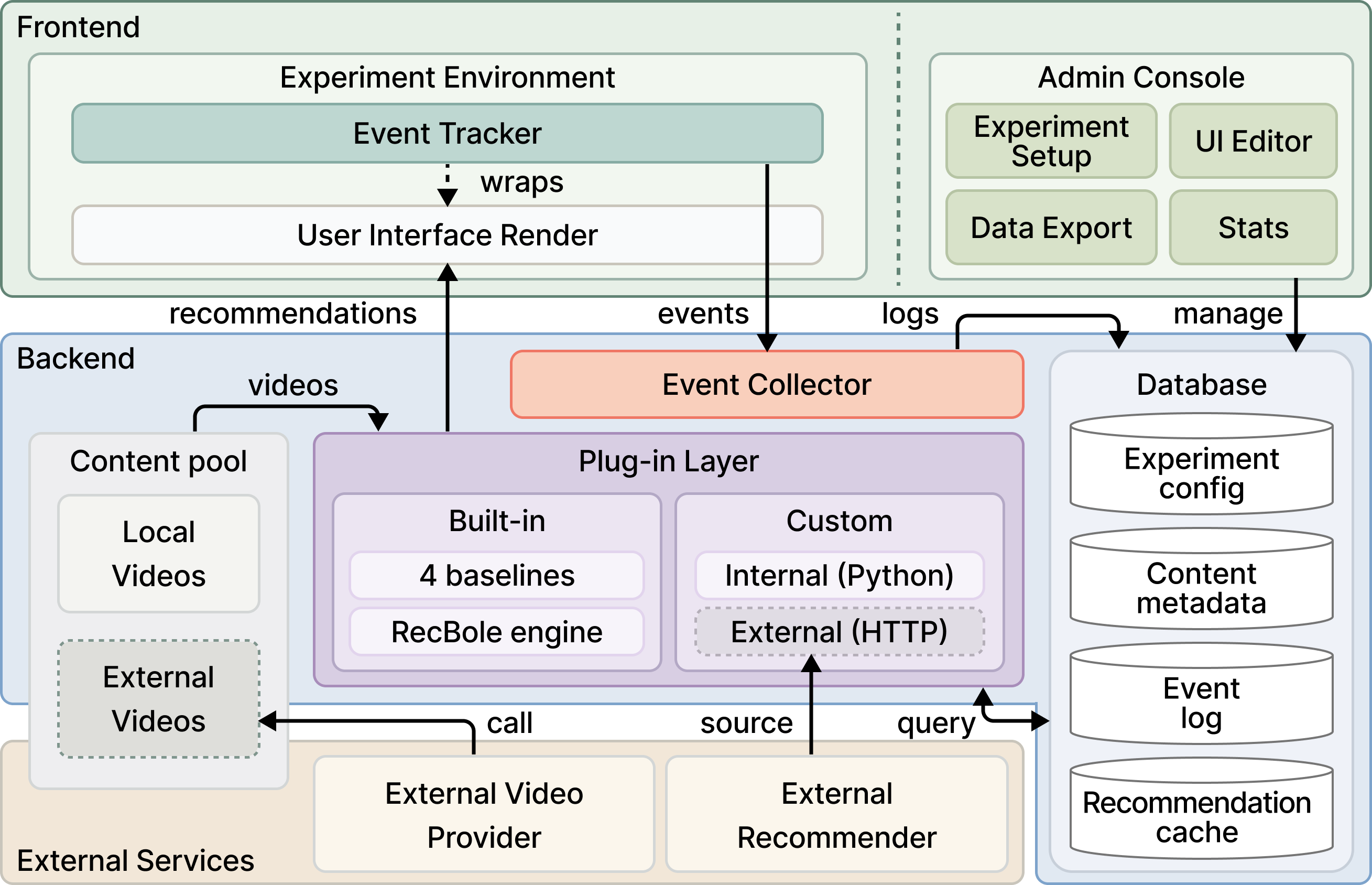}
\Description{Block diagram of the WatchLens architecture. A frontend experiment environment and admin console communicate with a backend that contains an event collector, a content pool, a plug-in layer hosting built-in and custom recommendation policies, and a database, with optional external video providers and external recommenders.}
\vspace{-5pt}
\caption{\proposed platform overview. Each user interface is wrapped by an Event Tracker that emits a standardized event stream with exposure context. Recommendation policies are dispatched through the Plug-in Layer, allowing the interface, feed policy, and watch-page policy to be configured independently per UserGroup over a shared content pool.}
\label{fig:architecture}
\vspace{-0.5cm}
\end{figure}

\begin{figure*}[t]
    \vspace{-0.2cm}
    \centering
    \includegraphics[width=0.99\textwidth]{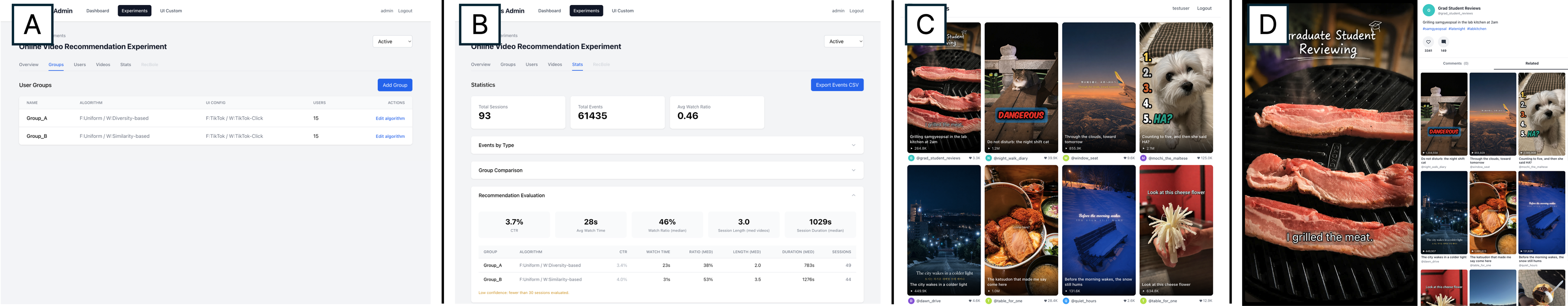}
    \Description{Four screenshots of WatchLens: the researcher-facing admin console showing the Groups tab and the Stats tab, and the participant-facing TikTok-preset interface showing the feed page and the watch page with related video recommendations.}
    \vspace{-0.2cm}
    \caption{\proposed. (A) and (B) show the researcher-facing admin console: (A) the Groups tab for configuring per-group policies, and (B) the Stats tab with automatically computed metrics. (C) and (D) show the participant-facing interface using the TikTok preset: (C) the feed page and (D) the watch page with related video recommendations beside the playing video.}
    \label{fig:overview}
    \vspace{-0.4cm}
\end{figure*}

\subsection{Platform Overview}
\label{sec:arch}

\proposed is designed so that researchers can vary only the factors they wish to compare in an online video recommendation experiment, while the rest of the system is reused without modification. 
To this end, the system is organized into three components (Figure~\ref{fig:architecture}): a frontend, a backend, and a plug-in layer inside the backend.
The \emph{frontend}, a React/TypeScript web application, presents content to participants and forwards every interaction to the backend as a standardized event stream. 
The \emph{backend}, built on FastAPI and PostgreSQL, persists experiment configurations and the event stream through an Event Collector, and exposes the data through analysis endpoints.
Recommendation policies reside in this isolated \emph{plug-in layer}, conforming to a single calling interface so that policies can be added or swapped without changes to the frontend or the logging layer. 
The entire system is packaged as a Docker Compose deployment for reproducible single-server setup.

\subsection{Modular Experiment Configuration}
\label{sec:config}

\vspace{-2pt}
\proposed defines a user study as an \textit{Experiment}, where each participant group to be compared is represented as a \textit{UserGroup}. UserGroups within the same Experiment share a content pool but each carries its own interface and recommendation policy configuration. 
By assigning different interfaces or policies to two UserGroups, researchers can directly compare the effect of that difference while holding the content pool and remaining configuration constant.

\vspace{-2pt}
\smallsection{User Interfaces}
\proposed ships with two built-in presets that mirror current video platform interfaces. 
The \textit{YouTube} preset implements a long-form interface with a 16:9 grid feed and a sidebar-based watch page, while the \textit{TikTok} preset implements a short-form interface with a 9:16 vertical feed and swipe-based video transitions. 
Beyond these presets, researchers can author their own interfaces through a \emph{custom UI editor} that supports a \textit{visual mode}, in which interfaces are assembled graphically from predefined building blocks, and a \textit{code mode}, in which interfaces are written as React components and compiled at runtime so that researchers iterate without a separate build step. 
Interfaces authored through either mode are built on top of the same set of tracking components (Section 3.3), so researchers can introduce new interfaces without writing event-instrumentation code or modifying the backend. 
All interfaces also adapt to desktop, tablet, and mobile browsers.

\vspace{-2pt}
\smallsection{Recommendation policies}
Each UserGroup can assign different policies to the feed and the watch page independently. Each policy is added as a plug-in to this layer, allowing built-in policies and custom researcher implementations to share the same integration path.
\proposed ships with four built-in baseline policies (random, popularity, recency, and similarity) and a learning-based policy via the built-in RecBole framework~\cite{zhao2021recbole, zhao2022recbole}, automatically trained on logged interactions for reproducible algorithm comparison. 
Custom policies can be added as in-process Python code for tight integration, or as external HTTP services for language-agnostic deployment.

\vspace{-2pt}
\smallsection{Content pool}
The content pool of each Experiment is curated by the researcher to match the research question at hand. 
Videos can be either hosted by the platform or embedded from external platforms (e.g., YouTube) within the same pool, allowing researchers to compose pools without constraints on domain or format.

\vspace{-5pt}
\subsection{Logging and Metrics}
\label{sec:logging}

The logging system of \proposed is built on two design decisions.

\textbf{First, all interfaces emit events under the same schema.} Behavioral tracking is decoupled from the interface code and performed by \emph{Event Trackers} that wrap the feed page, watch page, and video unit.
Regardless of interface type, they produce a standardized event stream spanning six categories: session lifecycle, page navigation, \textit{playback} (play, pause, seek, end, progress), impression, user interaction, and browser environment state (full schema in the project repository). 
Because the event contract is fixed across interfaces, cross-group and cross-study comparisons require no ad-hoc post-processing, even across different interface designs.

\textbf{Second, the signals needed for video recommendation analysis are recorded as first-class fields}.
The event schema captures both viewing behavior (play, pause, seek, end, watch duration, and watch ratio) and the exposure context indicating which recommendation policy produced the impression and at what position. 
The exposure context is attached as each event is emitted, so the linkage between playback and the recommendation that produced it is recorded directly rather than inferred post-hoc (Listing~\ref{lst:event}).

Building on these logs, \proposed automatically computes a set of metrics commonly reported in video recommendation studies~\cite{lin2023tree, xue2022resact, zhan2022deconfounding, zhao2023uncovering}: CTR, total watch time, session count, and the median values of watch duration, watch ratio, and session length. 
These metrics are computed both per group and aggregated across groups and exposed through the admin console, with evaluated events and metric definitions applied uniformly across groups to ensure cross-group comparability. 
In addition, all events are exported in an analysis-ready format with the exposure-context fields attached, enabling researchers to derive additional video-native quantities whose definitions depend on the specific research question.

\begin{lstlisting}[
  caption={A single event record in \proposed\ (selected fields). Playback signals and exposure context are recorded in the same event. The full schema is in the repository.},
  label={lst:event},
  basicstyle=\ttfamily\footnotesize,
  frame=single,
  captionpos=b,
    framesep=0pt,       
  ]
 {
    "event_type": "VIDEO_ENDED",
    "video_id": "qkpVPeNdNso",
    "watch_ratio": 1.0, 
    "watch_duration": 24.89,
    "position_in_feed": 2,
    "algorithm_feed": "random",
    "algorithm_watch": "similarity",
    "payload": {"completionRate": 1.0},
    "client_timestamp": "2026-04-24T19:02:29Z"
  }
\end{lstlisting}
\vspace{-4pt}

\subsection{Experimentation Workflow}
\label{sec:workflow}

A user study with \proposed proceeds in three stages within a single environment: setup, run, and analysis. 
In \textbf{setup}, researchers use the admin console to create an \textit{Experiment} and register its content pool. 
They then configure each \textit{UserGroup} with an interface and a recommendation policy, and register participants. 
In \textbf{run}, each participant is shown the interface assigned to their group, all interactions are logged following the schema in Section~\ref{sec:logging}, and built-in surveys can be delivered before the study, after the study, or between sessions, each independently enabled per experiment.
In \textbf{analysis}, researchers monitor per-group metrics and per-user behavioral trajectories through the admin console during the run, and export the event log together with its exposure-context fields for downstream analysis and reproducible re-analysis.

\section{Case Study}
\label{sec:exp}
\begin{table}[t]
\centering
\small
\setlength{\tabcolsep}{4pt}
\renewcommand{\arraystretch}{0.9}
\caption{Comparison of main metrics under the two watch-page policies. 
Values are user-level medians across 30 participants 
($*$: p < 0.05, Wilcoxon signed-rank test)}
\vspace{-5pt}
\label{tab:main}
\begin{tabular}{lcc}
\toprule
\textbf{Metric} 
& {\textbf{Diversity-based}} 
& {\textbf{Similarity-based}} \\
\midrule
Watch duration (s)            & 34.0  & 44.8$^{*}$  \\
Cumulative watch time (s)     & 700.2 & 1078.5$^{*}$ \\
Session length                 & 23  & 25$^{*}$  \\
Single-video chain rate     & 0.51  & 0.34$^{*}$  \\
\bottomrule
\end{tabular}

\end{table}
\begin{figure}[!t]
\centering
\vspace{-0.2cm}
\includegraphics[width=0.95\linewidth]{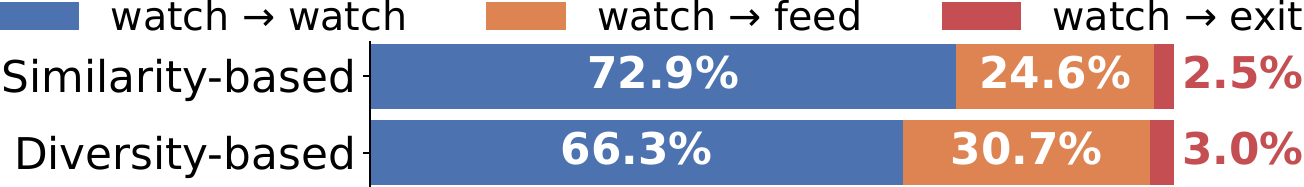}
\Description{Horizontal stacked bar chart of user transitions after leaving a watch page. Under the similarity-based policy, 72.9 percent of transitions continue to another watch page, 24.6 percent return to the feed, and 2.5 percent exit. Under the diversity-based policy, the corresponding values are 66.3, 30.7, and 3.0 percent.}
\vspace{-5pt}
\caption{Distribution of user transitions upon leaving a watch page, by policy. Averaged across participants.}
\label{fig:casestudy}
\vspace{-0.5cm}
\end{figure}

This case study demonstrates the applicability of \proposed. On a video platform, the home feed and the per-video watch page play distinct roles in recommendation: the feed determines where a user enters, while the watch page determines how the resulting session unfolds. 
We conducted a user study addressing the following question: \textbf{when the feed policy, the content pool, and the interface are held constant, how does varying only the watch-page policy affect session length and continuation behavior?} 
Such a comparison requires independent control of the policies on the two surfaces and exposure-aware logging of each behavior.
\proposed supports both, and we apply it to a session-level comparison.

\smallsection{Setup} Thirty university students participated in the study.\footnote{The study protocol was approved by the Yonsei University Institutional Review Board, and all participants provided informed consent.}
Each participant used \proposed for at least 20 minutes per day over four days. 
All participants used the same interface, which was based on the \textit{TikTok} preset and modified through the visual editor in Section~\ref{sec:config} to replace automatic swipe-based transitions with click-based transitions to recommended videos on the watch page. 
This modification allowed policy effects to be observed through users' explicit selection behavior. 
The content pool consisted of 1{,}000 popular short-form videos from five distinct categories.

\smallsection{Conditions and procedure} The feed policy was fixed across all participants and time points: a policy that drew videos uniformly from the five categories. 
The manipulated dimension was the watch-page policy. 
We compared two policies contrasting similarity and diversity.
The similarity-based policy constructs the recommendation list from videos within the same category as the currently playing video, while the diversity-based policy constructs the list uniformly across the five categories.  
Participants were assigned to two groups balanced on baseline daily short-form viewing time collected via a pre-study survey: Group~A received the diversity-based policy on days~1--2 and the similarity-based policy on days~3--4, while Group~B received the reverse order. 
This within-subject counterbalanced design balances policy effects against individual differences and time-dependent adaptation or fatigue effects.

\smallsection{Measures}
A \textit{session} is defined as one continuous span between a user entering \proposed and leaving it; a \textit{watch chain} is a run of consecutive watch-page visits within a session without returning to the feed.
To capture policy effects at the video, chain, and session level, we measured the following four metrics. \textit{Watch duration} is the time spent on a single watch page, and \textit{cumulative watch time} is the total watch time within one session.
\textit{Session length} counts the number of videos watched within a session, and the \textit{single-video chain rate} is the proportion of chains with a single video.
We additionally measured, upon leaving a watch page, the distribution of the user's next action over three transitions: \textit{continuation} to another watch page, \textit{return-to-feed}, and \textit{exit} from \proposed. Per-user medians are compared with Wilcoxon signed-rank tests.

\smallsection{Results} Table~\ref{tab:main} reports the main metrics under the two watch-page policies.
All four metrics shifted consistently in the direction of greater session retention under the similarity-based policy, with the largest relative change observed for cumulative watch time.
This is consistent with the watch-page policy affecting not only how long a single video is watched but also whether the user transitions to a next video, with these effects accumulating at the session level.
Figure~\ref{fig:casestudy} shows the user's next action after a watch page under each policy, indicating that users return to the feed more frequently under the diversity-based policy, while, in our study, the rate of leaving the platform itself was comparable across the two policies.

\smallsection{System-level reflection} This case study exercised \proposed's design decisions: independent feed and watch policy configuration (Section~\ref{sec:config}) isolated a single surface effect, and exposure context recorded in every event (Section~\ref{sec:logging}) made surface-level transition analysis directly 
available without custom instrumentation.

\vspace{-5pt}

\section{Conclusion}
\label{sec:conclusion}
We presented WatchLens, an open-source video-native platform for online video recommendation experiments, and demonstrated its applicability through a case study. 
Its modular configuration and exposure-aware logging let researchers isolate experimental factors across the feed and the watch page while preserving full exposure context for analysis. By lowering the infrastructure barrier to such experiments, we hope WatchLens enables researchers without access to industrial platforms to investigate how recommendation policies and interfaces shape real viewing behavior. 
A current limitation is that the platform is packaged as a single-server deployment whose behavior under high-concurrency load has not yet been characterized. 
Future work includes extending the video-serving infrastructure to a distributed deployment, integrating participant recruitment services, providing a native mobile client for short-form viewing, and building a shared repository for publishing and reproducing interface designs and recommendation policies.

\begin{acks}
This work was supported by the IITP grants funded by the Korea government (MSIT) (No. RS-2020-II201361; RS-2026-25520654).
\end{acks}


\bibliographystyle{ACM-Reference-Format}
\bibliography{BIB/bibliography}

\appendix

\end{document}